\documentclass[aps,prc,tightenlines,showpacs,preprint,showkeys,%
amssymb,byrevtex,nofootinbib,superscriptaddress]{revtex4}  

\usepackage{epsfig,bm,dcolumn}
\newcommand{\fslash}[1]{\ooalign{\hfil/\hfil\crcr$#1$}}
\begin{document}



\title{QCD sum rules for the anti-charmed pentaquark  }


\author{Yasemin Sarac}%
\email{ysarac@metu.edu.tr}

\affiliation{Physics Department,Middle East Technical University,
06531 Ankara, Turkey}

\author{Hungchong Kim}%
\email{hungchon@postech.ac.kr}

\affiliation{Department of Physics, Pohang University of Science
and Technology, Pohang 790-784,Korea}

\author{Su Houng Lee}%
\email{suhoung@phya.yonsei.ac.kr}

\affiliation{Institute of Physics and Applied Physics, Yonsei
University, Seoul 120-749, Korea}



\begin{abstract}

We present a QCD sum rule analysis for the anti-charmed
pentaquark state with and without strangeness. While the sum rules
for most of the currents are either non-convergent or dominated by
the $DN$ continuum, the one for the non-strange pentaquark current
composed of two diquarks and an antiquark, is convergent and has a
structure consistent with a positive parity pentaquark state after
subtracting out the $DN$ continuum contribution.   Arguments are
presented on the similarity between the result of the present
analysis and that based on the constituent quark models, which
predict a more stable pentaquark states when the antiquark is
heavy.

\end{abstract}


\maketitle

\section{Introduction}

The observation of the $\Theta^+$ by the LEPS
collaboration\cite{Leps03} and its subsequent confirmation have
brought a lot of excitements in the field of hadronic
physics\cite{OTHERS}. On the other hand,  there are increasing
number of experiments reporting negative results. In particular,
the latest experiments at JLAB\cite{Esmith} find no signal from
the photoproduction process on a deuteron nor on a proton target,
from which the $\Theta^+$ was observed earlier by the SAPHIR
collaboration with lower statistics.  Although the present
experimental results are quite confusing and
frustrating\cite{hicks}, one can not afford to give up further
refined experimental search, because if a pentaquark is  found, it
will provide a major and unique testing ground for QCD dynamics at
low energy.

Another multiplet to search for as possible pentaquark states are
those with one heavy antiquark. The H1 collaboration at HERA has
recently reported on the finding of an anti-charmed pentaquark
$\Theta_c (3099)$ from the $D^* p$ invariance mass
spectrum\cite{Aktas:2004qf}.  Unfortunately other experiments
could not confirm the
finding\cite{Litvintsev:2004yw,Karshon:2004kt,Link:2005ti}. While
the experimental search for the heavy pentaquark is as confusing
as that for the light, theoretically, the heavy and light
pentaquarks stand on quite different grounds.    Cohen showed that
the original prediction for the mass of the $\Theta^+$ based on
the SU(3) Skyrme model\cite{DPP97} is not valid because collective
quantization of the model for the anti-decuplet states is
inconsistent in the large $N_c$ limit\cite{Cohen:2003yi}.   In
contrast, many theories consistently predicted a stable heavy
pentaquark state.   The pentaquark with one heavy anti-quark was
first studied in Ref.~\cite{GSR87,Lip87} in a quark model with
color spin interaction. Then it has been studied in quark models
with flavor spin interaction\cite{Stan98} and Skyrme models
\cite{RS93,OPM94}, and with the recent experiments, attracted
renewed interests~\cite{CBSC,P04dw,Stewart:2004pd}, some of which were
motivated by the diquark-diquark\cite{JW03} and
diquark-triquark\cite{KL03a} picture.  Such states also appear
naturally in a coupled channel approach\cite{lutz}, and in the
combined  large $N_c$ and heavy quark limit of
QCD\cite{Cohen:2005bx}.    If the heavy pentaquark state is stable
against strong decay, as was predicted in the $D$ meson bound
solition models\cite{RS93}, it could only be observed from the
weak decay of the virtual $D$ meson.  {}From a constituent quark
model picture based on the color spin
interaction~\cite{Jaffe77}, one expects a strong
diquark correlation, from which one could have a stable
diquark-diquark-antiquark\cite{JW03} or diquark
triquark\cite{Lip87} structure.  The question is whether such
strong diquark structure will survive other non-perturbative QCD
dynamics in a multiquark environment and produce a stable
pentaquark state.  Such questions are being intensively pursued in
quark model
approaches\cite{Hiyama:2005cf,Stancu:2005jv,Maltman:2004qe}.  In
particular, an important question at hand is whether the net
attraction from the  diquark correlations in the pentaquark
configuration is stronger than that from the corresponding diquark
and additional quark-antiquark correlation present when the
pentaquark separates into a nucleon and a meson state. Since the
correlation are inversely proportional to the constituent quark
masses involved, the attraction is expected to be more effective
for pentaquark state with heavy antiquark. Another
non-perturbative approach that can be used to answer such question
is the QCD sum rule method.

There have been several QCD sum rule calculations for the light
pentaquark
states\cite{Zhu03,MNNRL03,SDO03,Eidemuller:2004ra,Lee:2004dp,Eidemuller:2005jm,
Matheus:2004gx}.    The application to the heavy pentaquarks was
performed by two of us in a previous work\cite{hung04}, where we
used a pentaquark current composed of two diquarks and an
antiquark, and found the sum rule to be consistent with a stable
positive parity pentaquark state.  The similar approach has been
applied to the sum rules for $D_s (2317)$~\cite{Kim05}. In this
work, we extend the previous QCD sum rule calculation to
investigate the anti-charmed pentaquark state with and without
strangeness using two different currents for each case.  We find a
convergent Operator Product Expansion (OPE) only for the
non-strange heavy pentaquark sum rule obtained with an
interpolating field composed of two diquarks and one anti-charm
quark, that has been previously used by us~\cite{hung04}.  The
stability of non-strange heavy pentaquark is consistent with the
result based on the quark model with flavor spin
interaction\cite{Stancu:2005jv}.   We then refine the convergent
sum rule by explicitly including the $DN$ two-particle irreducible
contribution. The importance of subtracting out such two-particle
irreducible contribution has been emphasized in
Ref.~\cite{Kondo05,Lee04,Kwon05} for the light pentaquark state.
In fact, estimating the contribution from the lowest two-particle
irreducible contribution is equally important in lattice gauge
theory calculations~\cite{Sasaki03,CFKK03} to isolate the signal
for the pentaquark state from the low-lying continuum state.   We
find that for the non-strange heavy pentaquark sum rule, including
the $DN$ continuum contribution tends to shift the position of the
pentaquark state downwards. Given the negative experimental
signatures of the charmed pentaquark states above the threshold,
the present result suggests that the anti-charmed pentaquark
states might be bound as was predicted in $D$ meson bound soliton
models.

This paper is organized as follows. In Section II, we introduce
the interpolating field for the $\Theta_c$ and discuss the
dispersion relations that we will be using.  Section III gives the
phenomenological side and Section IV gives the OPE side. The QCD
sum rules for $\Theta_c$ and their analysis are given in Section
V.

\section{QCD sum rules}

\subsection{Interpolating field for $\bm{\Theta_c}$}

Let us introduce the following two interpolating field for
$\Theta_c$,
\begin{eqnarray}
\Theta_{c1} & = & \epsilon^{abc} ( u^T_a C \gamma_\mu u_b)
\gamma_5\gamma_\mu d_c (\bar{c}_di\gamma_5d_d)\ , \nonumber \\
\Theta_{c2} & = & \epsilon^{abk} (\epsilon^{aef} u^T_e C \gamma_5
d_f)( \epsilon^{bgh} u^T_g C d_h)  C\bar{c}^T_k\ .
\label{currentc}
\end{eqnarray}
Here the Roman indices $a,b, \dots$ are color indices,  $C$
denotes charge conjugation, $T$ transpose.  Note that
$\Theta_{c1}$ is composed of a nucleon current ($proton$) and a pseudo
scalar current ($D$), while $\Theta_{c2}$ is composed of
diquark-diquark-antiquark and has been investigated in a previous
work\cite{hung04}.

For the charmed pentaquark with strangeness, we consider the
following two possible currents,
\begin{eqnarray}
\Theta_{cs1} & = & \epsilon^{abk} (\epsilon^{aef} u^T_e C  s_f)(
\epsilon^{bgh} u^T_g C d_h)  C\bar{c}^T_k\ , \nonumber \\
\Theta_{cs2} & = & \epsilon^{abk} (\epsilon^{aef} u^T_e C \gamma_5
s_f)( \epsilon^{bgh} u^T_g C d_h)  C\bar{c}^T_k\ .
\label{currentcs}
\end{eqnarray}
Here, instead of choosing $\Theta_{cs1}$ as a direct product of a
nucleon and a $D_s$ or a hyperon and a $D$ meson currents as in
$\Theta_{c1}$, we choose it to well represent a state having two
diquark structure with the same scalar quantum number but with
different flavor. Such configuration allows all the five
constituent quarks to be in the $s$-wave states, which will have the
lowest orbital energy and consequently could be the dominant
ground state configuration\cite{Stewart:2004pd}. Moreover, as we
will see, $\Theta_{c1}$ couples dominantly to the nucleon and $D$
meson state, suggesting that currents composed of a direct product
of a nucleon and a meson currents are not suitable for
investigating the properties of the pentaquark state.

Under parity transformation $q'(x')=\gamma_0 q(x)$, the $\Theta_c$
currents transform as,
\begin{eqnarray}
\Theta_{c1}'& =& - \gamma_0 \Theta_{c1} \ ,\quad
\Theta_{c2}' = \gamma_0  \Theta_{c2}\ , \nonumber \\
\Theta_{cs1}'& =& - \gamma_0 \Theta_{cs1} \ ,\quad
\Theta_{cs2}' = \gamma_0  \Theta_{cs2} . \label{qtran}
\end{eqnarray}

\subsection{Dispersion relation}

The first type of QCD sum rules for the heavy pentaquarks that we
will be using are constructed from the following time ordered
correlation function,
\begin{eqnarray}
\Pi_T(q)=i \int d^4 x e^{iq\cdot x} \langle 0 | T [\Theta_c (x),
{\bar \Theta_c }(0)] |0 \rangle \ \equiv \Pi_1(q^2)+\fslash{q}
\Pi_q(q^2)\  , \label{corr}
\end{eqnarray}
where $\Theta_c$ can be any of the currents in  Eq.~(\ref{currentc})
or in Eq.~(\ref{currentcs}), and $\Pi_q$, $\Pi_1$
are called the chiral even and chiral odd parts respectively.  As can
be seen in Eq. (\ref{qtran}), the currents are not eigenstates of
the parity transformation and can couple to both positive and
negative parity states.   The  spectral densities calculated from
the OPE of Eq.(\ref{corr}) are
matched to that obtained from the phenomenological assumption  in
the Borel-weighted dispersion integral,
\begin{eqnarray}
\int^{S_0}_{m_c^2} dq^2 e^{-q^2/M^2}W(q^2) {1\over \pi}{\rm Im} [
\Pi_i^{\rm phen} (q^2) - \Pi_i^{\rm ope} (q^2)] =0\ , \qquad
(i=1,q)\ , \label{sumrule1}
\end{eqnarray}
where $M^2$ is the Borel mass. Here, higher resonance
contributions are subtracted according to the QCD duality
assumption, which introduces the continuum threshold $S_0$.  We
have also introduced an additional weight function $W(q^2)$
for later use.

In this work, we will also work with the ``old-fashioned''
correlation function, which is defined as\cite{SDO03}
\begin{eqnarray}
\Pi_T(q)=i \int d^4 x e^{iq\cdot x} \langle 0 | \theta(x^0)
\Theta_c(x) \bar{\Theta}_c (0) |0 \rangle\ . \label{corr2}
\end{eqnarray}
This type of  correlation function has been used in
projecting out positive and negative parity nucleon
states~\cite{Jido:1996ia}.  We then divide the imaginary part into
the following two parts, which are defined only for $q_0>0$,
\begin{eqnarray}
\frac{1}{\pi} {\rm Im} \Pi(q_0) & = & A(q_0) \gamma^0+B(q_0)\ .
\end{eqnarray}
One should note that these can be identified with the imaginary
part calculated from Eq.(\ref{corr}),
\begin{eqnarray}
A(q_0) & = & \frac{1}{\pi} {\rm Im} \Pi_q(q_0) q_0 \nonumber \\
B(q_0) & = & \frac{1}{\pi} {\rm Im} \Pi_1(q_0),
\end{eqnarray}
for $q_0>0$.

Now, depending on the parity of the current $\Theta_c$ in
Eq.(\ref{qtran}), one can extract the positive or negative-parity
physical state only by either adding or subtracting $A$ and $B$.
That is,  the spectral density for the positive and negative
parity physical states will be as follows,
\begin{eqnarray}
\rho^{\pm}(q_0)=\bigg\{
\begin{array}{cc}
A(q_0)\mp B(q_0) &  {\rm For} \,\,\, \Theta_{c1},\Theta_{cs1}  \\
A(q_0)\pm B(q_0) &  {\rm For} \,\,\, \Theta_{c2},\Theta_{cs2}
\end{array}.
\label{rhopm}
\end{eqnarray}
The sum rules are then obtained by again matching the corresponding
spectral density from the OPE and phenomenological side,
\begin{eqnarray}
\int^\infty_0 dq_0 e^{-q_0^2/M^2 }
[\rho_{\rm phen}^\pm(q_0)-\rho_{\rm ope}^\pm (q_0)]
 = 0\ . \label{sumrule2}
\end{eqnarray}

\section{Phenomenological side}

\subsection{$\Theta_{c1}, \Theta_{cs1}$}

For $\Theta_{c1}$ current, the interpolating field couples to
a positive parity state as,
\begin{eqnarray}
\langle 0 | \Theta_{c1} (x) | \Theta_c ({\bf p}) :P = + \rangle =
\lambda_{+,c1}~ \gamma_5 U_\Theta ({\bf p}) e^{-i p\cdot x}\ ,
\end{eqnarray}
and to a negative parity state as,
\begin{eqnarray}
\langle 0 | \Theta_{c1} (x) | \Theta_c ({\bf p}) :P= - \rangle  =
\lambda_{-,c1} U_\Theta ({\bf p}) e^{-i p\cdot x}\ .
\end{eqnarray}
Here, $\lambda_{\pm,c1}$ denotes the coupling strength
between the
interpolating field and the physical state with the specified
parity.  Similar relations will hold for $\Theta_{cs1}$.  Using
these, we obtain the phenomenological side of Eq.~(\ref{corr})
separated into chiral even ($\Pi_q$) and odd ($\Pi_1$) parts,
which are defined to be the parts proportional to $\fslash{q}$ and
$1$ respectively.

As was first pointed out in Ref.\cite{Kondo05}, the correlation
function can also couple to the $DN$ continuum state, whose
threshold could be lower than the expected $\Theta_c$ mass.  Its
phenomenological contribution can be estimated by using,
\begin{eqnarray}
\langle 0 | \Theta_{c1} | DN({\bf p})  \rangle = i \lambda_{DN,c1}
U_N ({\bf p})\ . \label{lambda1}
\end{eqnarray}
Combining these two contributions, we find
\begin{eqnarray}
\Pi^{\rm phen}_{T,c1} (q) = - |\lambda_{\pm,c1}|^2 ~ { \fslash{q}
\mp m_{\Theta} \over q^2 -m^2_{\Theta} } -i|\lambda_{DN,c1}|^2\int
d^4p\frac{(\fslash{p}+m_N)}{p^2-m_N^2}\frac{1}{(p-q)^2-m_D^2}+\cdot
\cdot \cdot ,
\end{eqnarray}
where the minus (plus) sign in front of $m_{\Theta}$ is for
positive (negative) parity. The dots denote higher resonance
contributions that should be parameterized according to QCD
duality. It should be noted however that higher resonances with
different parities contribute differently to the chiral-even and
chiral odd parts \cite{Jin:1997pb}. Thus, $\Pi^{\rm phen}_q$ and
$\Pi^{\rm phen}_1$ constitute  separate sum rules.  For
$\Theta_{cs1}$, the $D$ meson should be replaced by the $D_s$ meson.

The corresponding spectral density for the pole and $DN$
contributions are given respectively by
\begin{eqnarray}
{1\over \pi} {\rm Im} \Pi^{\rm pole}_{T,c1} (q) &=& \fslash{q}
|\lambda_{\pm,c1}|^2  \delta (q^2 -m^2_{\Theta})
      \mp m_\Theta |\lambda_{\pm,c1}|^2 \delta (q^2 -m^2_{\Theta}),
   \nonumber   \\
{1\over \pi} {\rm Im} \Pi^{\rm DN}_{T,c1} (q) &=&
\fslash{q}|\lambda_{DN,c1}|^2\frac{q^2+m_N^2-m_D^2}{32\,\pi^2\,
q^4}\sqrt{q^4-2q^2(m_N^2+m_D^2)+(m_N^2-m_D^2)^2}
\nonumber \\
&&+|\lambda_{DN,c1}|^2\frac{2 m_N}{32\, \pi^2\,
q^2}\sqrt{q^4-2q^2(m_N^2+m_D^2)+(m_N^2-m_D^2)^2}. \label{phen_dn1}
\end{eqnarray}
We notice that the chiral-odd part  has opposite sign depending on
the parity while the chiral even part has positive-definite
coefficient.

\subsection{$\Theta_{c2}, \Theta_{cs2}$}

As can be seen in Eq.(\ref{qtran}), $\Theta_{c2}$ transforms
differently compared to $\Theta_{c1}$ under parity. Thus, the
couplings to the interpolating field are
\begin{eqnarray}
\langle 0 | \Theta_{c2} (x) | \Theta_c ({\bf p}) :P = + \rangle &
= &  \lambda_{+,c2}  U_\Theta ({\bf p}) e^{-i p\cdot x}\ ,
\nonumber \\
\langle 0 | \Theta_{c2} (x) | \Theta_c ({\bf p}) :P= - \rangle & =
& \lambda_{-,c2}~ \gamma_5 U_\Theta ({\bf p}) e^{-i p\cdot x}\ .
\end{eqnarray}
Similarly, the coupling to the $DN$ continuum state changes as
follows,
\begin{eqnarray}
 \langle 0 | \Theta_{c2} | DN({\bf p})  \rangle = \lambda_{DN,c2} \gamma_5
 U_N ({\bf p}). \label{lambda2}
\end{eqnarray}
Combining these changes, we find,
\begin{eqnarray}
\Pi^{\rm phen}_{T,c2} (q) = - |\lambda_{\pm,c2}|^2 ~ { \fslash{q}
\pm m_{\Theta} \over q^2 -m^2_{\Theta} } +i|\lambda_{DN,c2}|^2\int
d^4p\frac{\gamma_5(\fslash{p}+m_N)\gamma_5}{p^2-m_N^2}\frac{1}{(p-q)^2-m_D^2}+\cdot
\cdot \cdot .
\end{eqnarray}
Consequently, the  spectral densities are,
\begin{eqnarray}
{1\over \pi} {\rm Im} \Pi^{\rm pole}_{T,c2} (q) &=& \fslash{q}
|\lambda_{\pm,c2}|^2  \delta (q^2 -m^2_{\Theta})
      \pm m_\Theta |\lambda_{\pm,c2}|^2 \delta (q^2 -m^2_{\Theta}),
\nonumber      \\
{1\over \pi} {\rm Im} \Pi^{\rm DN}_{T,c2} (q) &=&
\fslash{q}|\lambda_{DN,c2}|^2\frac{q^2+m_N^2-m_D^2}{32\,\pi^2\,
q^4}\sqrt{q^4-2q^2(m_N^2+m_D^2)+(m_N^2-m_D^2)^2}
\nonumber \\
&&-|\lambda_{DN,c2}|^2\frac{2 m_N}{32\, \pi^2\,
q^2}\sqrt{q^4-2q^2(m_N^2+m_D^2)+(m_N^2-m_D^2)^2}. \label{phen_dn2}
\end{eqnarray}

\subsection{Phenomenological side}

The final form for the phenomenological side to be used in
Eq.(\ref{sumrule2}) can be obtained from combining
Eq.(\ref{phen_dn1}) or Eq.(\ref{phen_dn2}) according to
Eq.(\ref{rhopm}), both of which are given in the following form,
\begin{eqnarray}
\rho_{\rm phen}^\pm(q_0)=|\lambda_{\pm}|^2 \delta(q_0-m_{\Theta})
+ \theta(\sqrt{s_0}-q_0) \rho^\pm_{DN}(q_0) +\theta(q_0
-\sqrt{s_0}) \rho^\pm_{\rm cont}(q_0)\ , \label{phen-rho}
\end{eqnarray}
where the usual duality assumption has been used to represent the
higher resonance contribution above the continuum threshold
$\sqrt{s_0}$; i.e.,
$\rho^\pm_{\rm cont}(q_0) =\rho^\pm_{\rm ope}(q_0)$.
The spectral density for the two-particle irreducible part
is given by
\begin{eqnarray}
\rho_{DN,c1}^\pm(q_0)& = & \frac{|\lambda_{DN,c1}|^2}{32 \pi^2 }
\sqrt{ (q_0-m_D)^2-m_N^2 } \sqrt{(q_0+m_D)^2-m_N^2 } \nonumber
\\ && \times
\frac {(q_0\pm m_N)^2-m_D^2} {q_0^3}\ ,  \\
\rho_{DN,c2}^\pm(q_0)& = & \frac{|\lambda_{DN,c2}|^2}{32 \pi^2 }
\sqrt{ (q_0-m_D)^2-m_N^2 } \sqrt{(q_0+m_D)^2-m_N^2 } \nonumber
\\ && \times
\frac {(q_0\mp m_N)^2-m_D^2} {q_0^3}\ .
\end{eqnarray}
We substitute the above into the Borel transformed
dispersion relation in Eq.(\ref{sumrule2}).

\section{OPE side}

\subsection{ $\Theta_{c1}$}

Here, we present the
result for $\Theta_{c1}$. To keep the charm quark mass finite, we
use the momentum-space expression for the charm quark propagators.
For the light quark part of the correlation function, we calculate
in the coordinate-space, which is then Fourier-transformed to the
momentum space in $D$-dimension. The resulting light-quark part is
combined with the charm-quark part before it is dimensionally
regularized at $D=4$.

\begin{figure}
\centering \epsfig{file=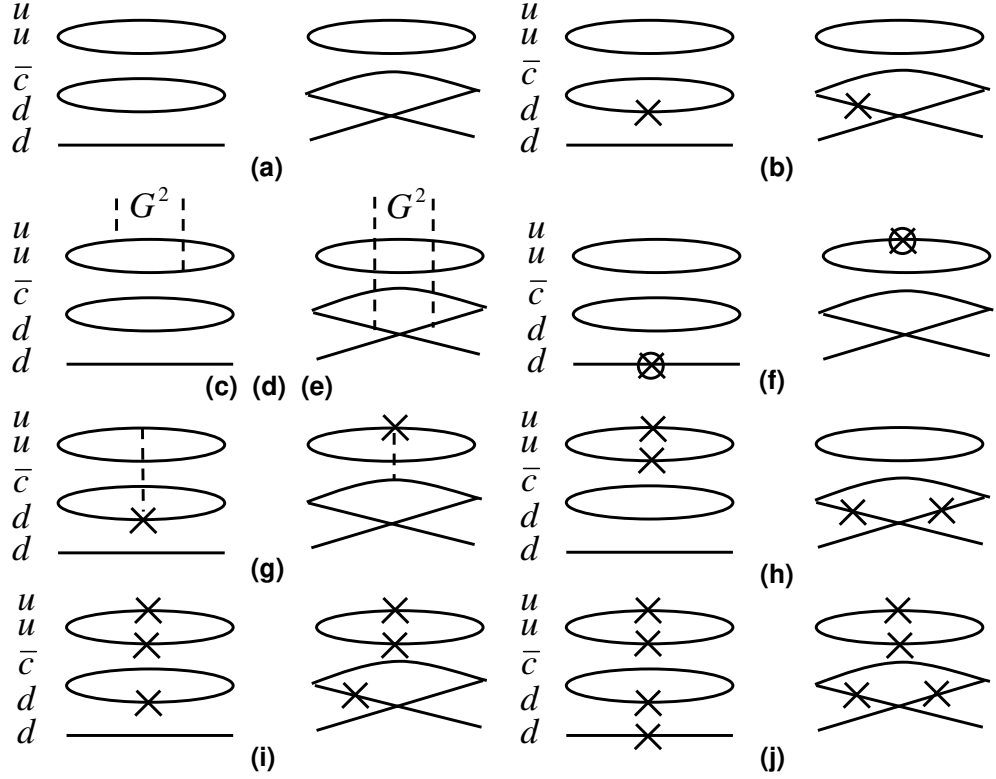, width=1.\hsize}
\caption{Schematic OPE diagrams for the current $\Theta_{c1}$ in
Eq.(\ref{currentc}). Each
label corresponds to that in Eq.(\ref{ope-c1}).
 The solid lines denote quark (or anti-charm
quark) propagators and the dashed lines are for gluon. The crosses
denote the quark condensate, and the crosses with circle represent
the mixed quark gluon condensates.  (c) represents diagrams
proportional to gluon condensate with gluons lines attached to the
light quarks only, (d) represents those where the gluons are
attached to the heavy quarks only, while (e) represents those
where one gluon is attached to the heavy quark and the other to a
light quark in all possible ways. (f) and (g) represent all
diagrams that contain the quark-gluon condensate. } \label{fig1}
\end{figure}

Our OPE is given by
\begin{eqnarray}
\Pi^{\rm ope,c1} (q)&=&  \Pi^{(a)}+\Pi^{(b)}+\Pi^{(c)}
                     +\Pi^{(d)}+\Pi^{(e)}
\nonumber\\
                  &&
+ \Pi^{(f)}+\Pi^{(g)}+\Pi^{(h)}+\Pi^{(i)}+\Pi^{(j)},
\end{eqnarray}
where the superscript indicates each diagram
in Fig.~\ref{fig1}. The imaginary part
of each diagram is calculated as
\begin{eqnarray}
{1\over \pi} {\rm Im} \Pi^{(a)} (q^2) &=& \frac{11}{ 5! \, 6!
\,2^{13} \, \pi^8} \int_{0}^{\Lambda} du \, \frac{1}{(1-u)^{5}}
\left\{ \fslash{q} \left[ - 36 u (1-u) [-L(u)]^5 \right. \right.
\nonumber\\
&&+ \left. \left. 120 q^2 u^2 (1-u)^2 [-L(u)]^4 \right] + m_c72
[-L(u)]^5 \right\}
\ ,\nonumber \\
{1\over \pi} {\rm Im} \Pi^{(b)} (q^2) &=& {5
\langle\bar{q}q\rangle\over4!4!2^{9}\pi^8}\int_{0}^{\Lambda}du
\frac{1}{(1-u)^3}\left\{\fslash{q}16m_cu[-L(u)]^3\right. \nonumber
\\&& +\left.8q^2u(1-u)[-L(u)]^3-[-L(u)]^4\right\}
\ ,\nonumber \\
{1\over \pi} {\rm Im} \Pi^{(c)} (q^2) &=&
-\frac{11\langle\frac{\alpha_s}{\pi}G^2\rangle}{3\cdot4!2^{14}\pi^6}
\int_{0}^{\Lambda}du\frac{1}{(1-u)^3}\Bigg\{\fslash{q}\Big[-3u(1-u)[-L(u)]^3
\nonumber\\
&&+6q^2u^2(1-u)^2[-L(u)]^2-\frac{8}{11}(1-u)[-L(u)]^3\Big]
\nonumber\\
&&+\frac{12}{11}m_c[-L(u)]^3 \Bigg\}
\ ,\nonumber\\
{1\over \pi} {\rm Im} \Pi^{(d)} (q^2) &=&
\frac{11\langle\frac{\alpha_s}{\pi}G^2\rangle}{3!6!2^{12}\pi^6}
\int_{0}^{\Lambda}du\frac{u^3}{(1-u)^5}
\Bigg\{\fslash{q}m_c^2\Big[-3u(1-u)[-L(u)]^2
\nonumber\\
&&+4q^2u^2(1-u)^2[-L(u)]\Big ]+m_c\Big[\frac{4}{11}[-L(u)]^3
\nonumber\\
&& +\frac{6}{11}q^2(1-u)^2[-L(u)]^2\Big]\Bigg\} \
,\nonumber\\
{1\over \pi} {\rm Im} \Pi^{(e)} (q^2) &=&
\frac{\langle\frac{\alpha_s}{\pi}G^2\rangle}{4!5!3\cdot2^{10}\pi^6}
\int_{0}^{\Lambda}du\frac{u}{(1-u)^4}
\Bigg\{\fslash{q}\Big[\Big(96u(1-u)+5(1-u)\Big)[-L(u)]^3
\nonumber\\
&&-192q^2u^2(1-u)^2[-L(u)]^2\Big ]+90m_c[-L(u)]^3 \Bigg\} \
,\nonumber\\
{1\over \pi} {\rm Im} \Pi^{(f)}(q^2)&=&
\frac{5\langle\bar{q}g\sigma\cdot
Gq\rangle}{3!3!2^{11}\pi^6}\int_{0}^{\Lambda}du
\frac{1}{(1-u)^2}\Big\{12\fslash{q}m_cu[-L(u)]^2
\nonumber\\
&&-[-L(u)]^3+6q^2u(1-u)[-L(u)]^2\Big\}
\ ,\nonumber\\
{1\over \pi} {\rm Im} \Pi^{(g)} (q^2) &=&
\frac{\langle\bar{q}g\sigma\cdot Gq\rangle}{3!4!2^ {10}\pi^6}
\int_{0}^{\Lambda}du\frac{1}{(1-u)^3}
\Bigg\{\fslash{q}m_c\Big[12u(1-u)[-L(u)]^2
\nonumber\\
&&-60u^2[-L(u)]^2\Big]-12(1-u)[-L(u)]^3+72q^2u(1-u)^2[-L(u)]^2
\nonumber\\
&&-\frac{u}{2}[-L(u)]^3+3q^2u^2(1-u)[-L(u)]^2\Bigg\}
 ,\nonumber\\
{1\over \pi} {\rm Im} \Pi^{(h)} (q^2) &=& \frac{ \langle \bar {q}
q \rangle^2}{9\cdot2^{9}\pi^4}
\int_{0}^{\Lambda}du\frac{1}{(1-u)^2} \Bigg\{\fslash{q}\Big[
12u(1-u)[-L(u)]^2 \nonumber
\\&&-16q^2u^2(1-u)^2[-L(u)]+3(1-u)[-L(u)]^2\Big ]+27 m_c[-L(u)]^2\Bigg \}
\ ,\nonumber \\
{1\over \pi} {\rm Im} \Pi^{(i)} (q^2) &=&
\frac{5\langle\bar{q}q\rangle^3}{9\cdot2^{4}\pi^2}\int_{0}^{\Lambda}du
\left\{-\fslash{q}m_cu+[-L(u)]-2q^2u(1-u)\right\}
\ ,\nonumber \\
{1\over \pi} {\rm Im} \Pi^{(j)} (q^2) &=&
\frac{\langle\bar{q}q\rangle^4}{216}\left(-\fslash{q}+22m_c\right)
\delta(q^2-m_c^2). \label{ope-c1}
\end{eqnarray}
Here the upper limit of the integrations is given by
$\Lambda=1-m_c^2 /q^2$ and $L(u)=q^2u(1-u)-m_c^2u$. Our OPE
calculation has been performed up to dimension 12 here. Up to
dimension 5, we include all the gluonic contributions represented
by the gluon condensate and the quark-gluon mixed condensate.
Beyond the dimension 5, we have included only tree-graph
contributions which are expected to be important among higher
dimensional operators.    Other diagrams containing gluon
components are expected to be suppressed by the small QCD
coupling.  Therefore, the higher order tree-graphs, which are the
higher order quark condensates, will be able to give us an
estimate on how big the typical higher order corrections should be
beyond dimension 5.   The integrations can be done analytically
but we skip the messy analytic expressions. For the charm-quark
propagators with two gluons attached, we use the momentum-space
expressions given in Ref.~\cite{Reinders:1984sr}. The Wilson
coefficients for light-quark condensates come from $\langle
\bar{q} q\rangle^n $, where $n=2,3,4$.  This is in contrast with
the OPE for $\Theta_{c2}$, where the Wilson coefficient are
non-zero only for $n=4$.

The first important question to ask in the OPE is whether it is
sensibly converging as an asymptotic expansion.   For that, we
choose to plot the Borel transformed OPE appearing in
Eq.(\ref{sumrule2}) after subtracting out the continuum
contribution,
\begin{eqnarray}
\Pi^{(j)}(M^2)= \int^\infty_0 dq_0~e^{-q_0^2/M^2 }
[\rho_{\rm ope}^{\pm,(j)} (q_0)-\rho_{\rm cont}^{\pm,(j)}(q_0)]
 = 0.  \label{borel-opec1}
\end{eqnarray}
Here $j=a,b,c..$ denotes each contribution in the OPE in
Eq.(\ref{ope-c1}) after adding according to the rules in
Eq.(\ref{rhopm}).

\begin{figure}
\centering \epsfig{file=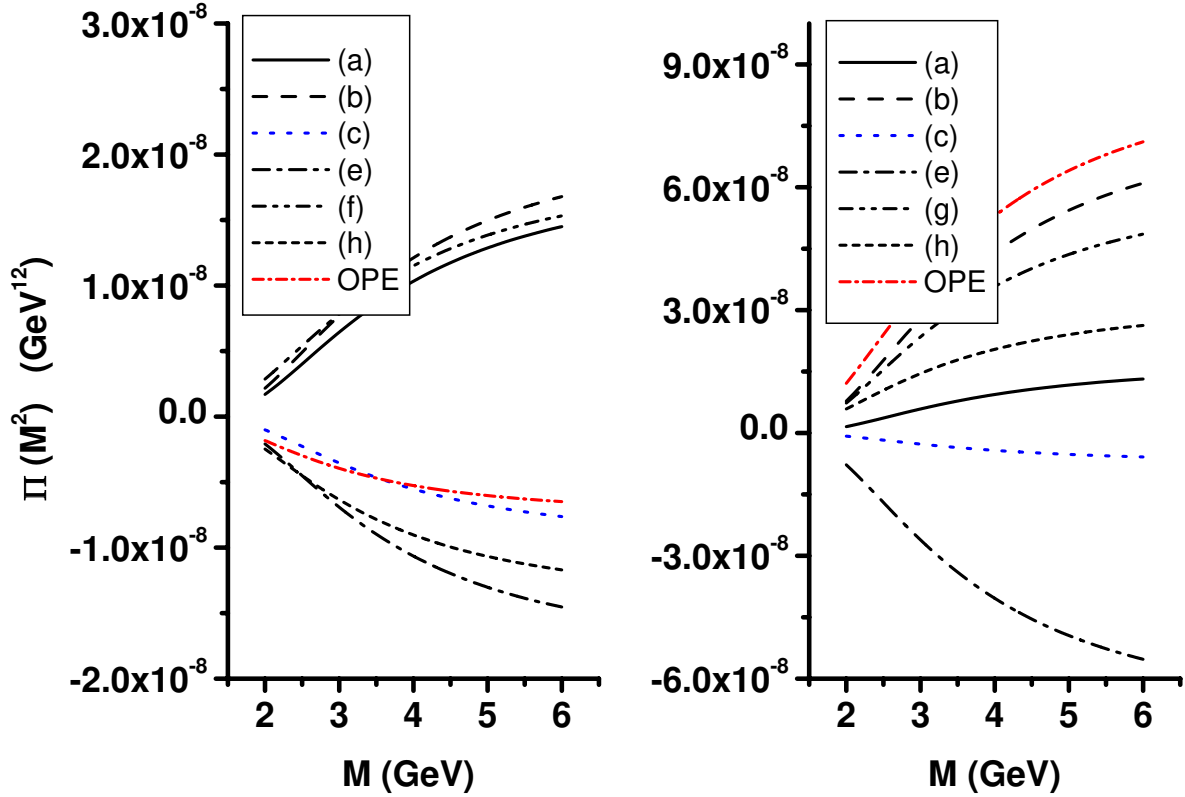, width=1.\hsize} \caption{OPE
as defined in Eq.(\ref{borel-opec1}) for the current $\Theta_{c1}$
and $S_0=(3.3~{\rm GeV})^2$.  The left (right) figure is for
positive (negative) parity case.  The solid line (a) represents the
perturbative contribution. The line specified as OPE
represents the sum of the power
corrections only. (c) represents the gluon condensates.  Other
labels represent contribution from each term in
Eq.(\ref{ope-c1}). Here we plot only a few selected terms in the OPE.}
\label{fig-borel-opec1}
\end{figure}

We use the following QCD parameters in our sum
rules~\cite{Shifman:bx,SDO03},
\begin{eqnarray}
m_s  =   0.12~{\rm GeV} \ &,& \quad m_c = 1.26~{\rm GeV} \ , \nonumber \\
\left \langle {\alpha_s \over \pi} G^2 \right \rangle = (0.33~{\rm
GeV})^4 \ &,& \quad \langle G^3 \rangle = 0.045 ~{\rm GeV}^6\ ,
\nonumber \\
\langle \bar {q} q \rangle = -(0.23~{\rm GeV})^3 \ &,& \quad
\langle \bar {s} s \rangle = 0.8 \langle \bar {q} q \rangle\ ,
\nonumber \\
\langle \bar {q}g \sigma \cdot G q \rangle  =  (0.8 ~{\rm GeV}^2) \times
\langle \bar {q} q \rangle  \ &,& \quad \langle \bar {s}g \sigma
\cdot G s \rangle  =  (0.8 ~{\rm GeV}^2) \times \langle \bar {s} s \rangle.
\end{eqnarray}
Fig. (\ref{fig-borel-opec1}) represents the OPE as defined in
Eq.(\ref{borel-opec1}) with the imaginary part in
Eq.(\ref{ope-c1}).  One notes that for the negative parity case,
the perturbative contribution is only a small fraction of the OPE,
and hence do not converge.  For the positive parity case, the
power corrections alternate in signs, and the gluon condensate,
which represents the light diquark correlation, is only a small
correction to the power correction.  Hence, such structure, would
hardly couple to a pentaquark state, and it is meaningless to
perform a detailed QCD sum rule analysis.
We present the result with the continuum threshold $S_0 =(3.3~ {\rm GeV})^2$.
This value is chosen in the range $\sqrt{S_0}=3.2 - 3.6 $ GeV,
which has been used to analyze the anticharmed-pentaquark sum rule
in Ref~\cite{hung04}.
However changing $S_0$ does not
change the relative strength of each contribution, and hence the
conclusion of this section.  We will therefore, analyze the
subsequent OPE with the same threshold.

\subsection{ $\Theta_{c2}$}

\begin{figure}
\centering \epsfig{file=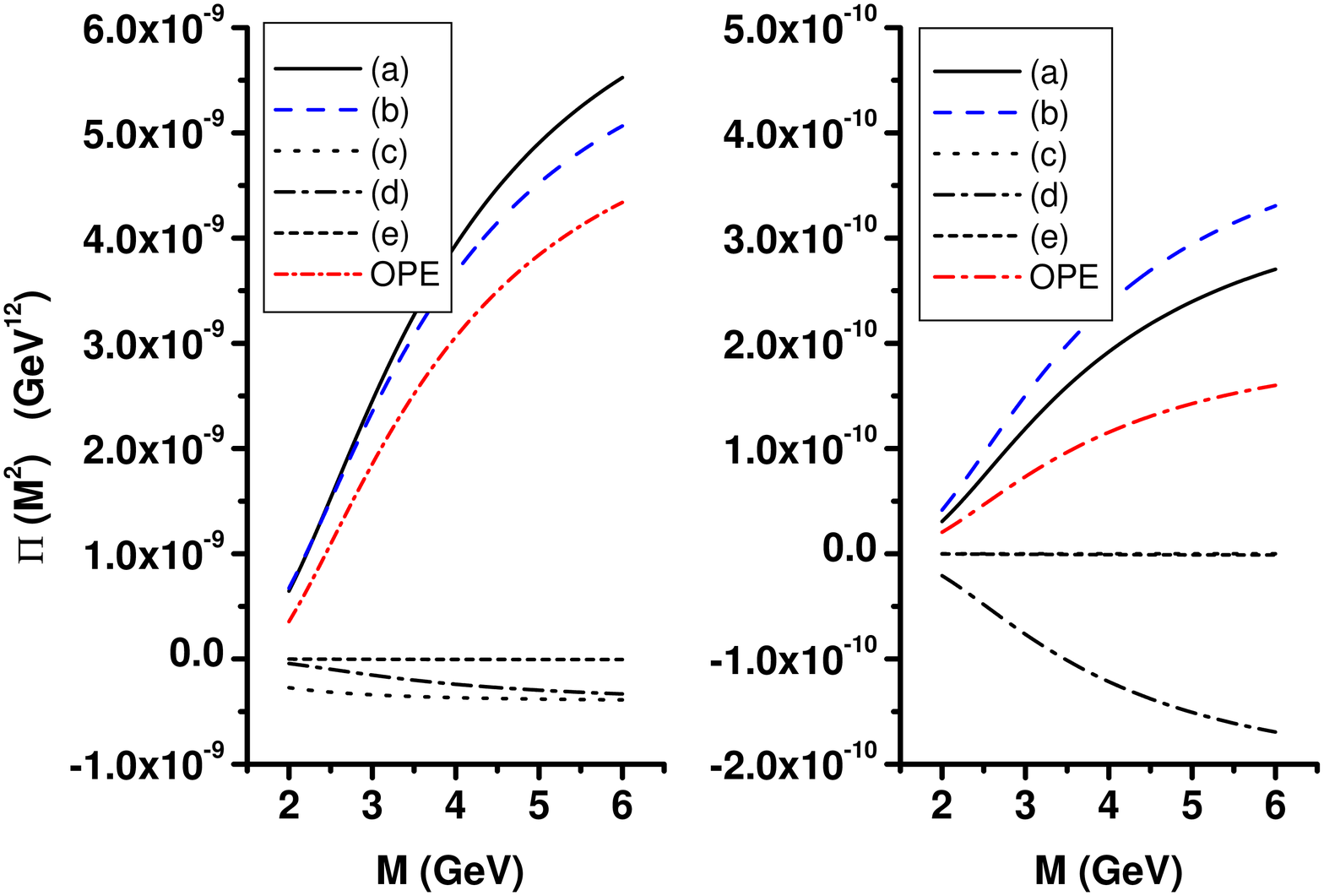, width=1.\hsize} \caption{
Similar figure as Fig.\ref{fig-borel-opec1} for the current $\Theta_{c2}$.
Here each label represents contribution from each term in Eq.(\ref{ope-c2}).
The gluon condensates (b) are the dominant power
correction in the positive parity channel (left figure).}
\label{fig-borel-opec2}
\end{figure}

The OPE for $\Theta_{c2}$ are given in Ref.\cite{hung04}. Here, we
rewrite the result for completeness,
\begin{eqnarray}
{1\over \pi} {\rm Im} \Pi^{(a)} (q^2) &=& -{1\over 5\cdot 5!\
2^{12} \pi^8} \int^{\Lambda}_0 du \frac{1}{(1-u)^5}
\left\{\fslash{q} (1-u) + m_c \right\} \ [-L(u)]^5
\ ,\nonumber \\
{1\over \pi} {\rm Im} \Pi^{(b)} (q^2) &=& -{1\over 3!\ 3!\ 2^{10}
\pi^6} \left \langle {\alpha_s \over \pi} G^2 \right \rangle
\int^{\Lambda}_0 du \frac{1}{(1-u)^3} \left\{ \fslash{q} (1-u) +
m_c \right\} [ -L(u)]^3
\ ,\nonumber \\
{1\over \pi} {\rm Im} \Pi^{(c)} (q^2) &=& -{1\over 54} \langle
\bar {q} q \rangle^4 (\fslash{q}  + m_c )~ \delta (q^2 - m_c^2)
\ ,\nonumber \\
{1\over \pi} {\rm Im} \Pi^{(d)} (q^2) &=& -{1\over 5!\ 3!\ 3\cdot
2^{10} \pi^6} \left \langle {\alpha_s \over \pi} G^2 \right
\rangle \int^{\Lambda}_0 du  \frac{u^3}{(1-u )^5}
\nonumber \\
&\times & \Bigg\{3 m_c^2 \fslash{q} (1-u) + m_c (1-u)(3-5u)q^2+2u
m_c^3 \Bigg\}  [ -L(u)]^2
\ ,\nonumber \\
{1\over \pi} {\rm Im} \Pi^{(e)} (q^2) &=& -{\left \langle G^3
\right \rangle \over 5!\ 4!\ 2^{13} \pi^8} \int^{\Lambda}_0 du
\frac{u}{(1-u)}~ \Bigg \{ \fslash{q} \left [ q^2 \left ( {5u \over
2}-1 \right ) (1-u)  -m_c^2 \left ( {3u \over 2} + 7 \right )
\right ]
\nonumber \\
&&+6 m_c q^2 (2u-1) -2 m_c^3 {3u+1\over 1-u}  \Bigg \} [-L(u)].
\label{ope-c2}
\end{eqnarray}
The diagrams corresponding to every term above, denoted by the
superscripts $(a)-(e)$, can be found in Ref.~\cite{hung04}.

Fig. (\ref{fig-borel-opec2}) represents the OPE as defined in
Eq.(\ref{borel-opec1}) with the imaginary part in
Eq.(\ref{ope-c2}).  As can be seen from the left figure, the OPE
without the perturbative contribution is dominated by the gluon
condensate coming from the light diquarks. This suggests that the
diquark correlation is the dominant interaction among the quarks
and heavy antiquark in the positive parity channel. Moreover, the
perturbative contribution is larger than sum of the power
corrections denoted as ``OPE'' in the figure. Therefore, the
pentaquark could couple strongly to this current and a detailed
QCD sum rule analysis is sensible.   The situation changes for the
negative channel, where the power corrections have alternating
signs, and hence becomes less reliable.

\begin{figure}
\centering \epsfig{file=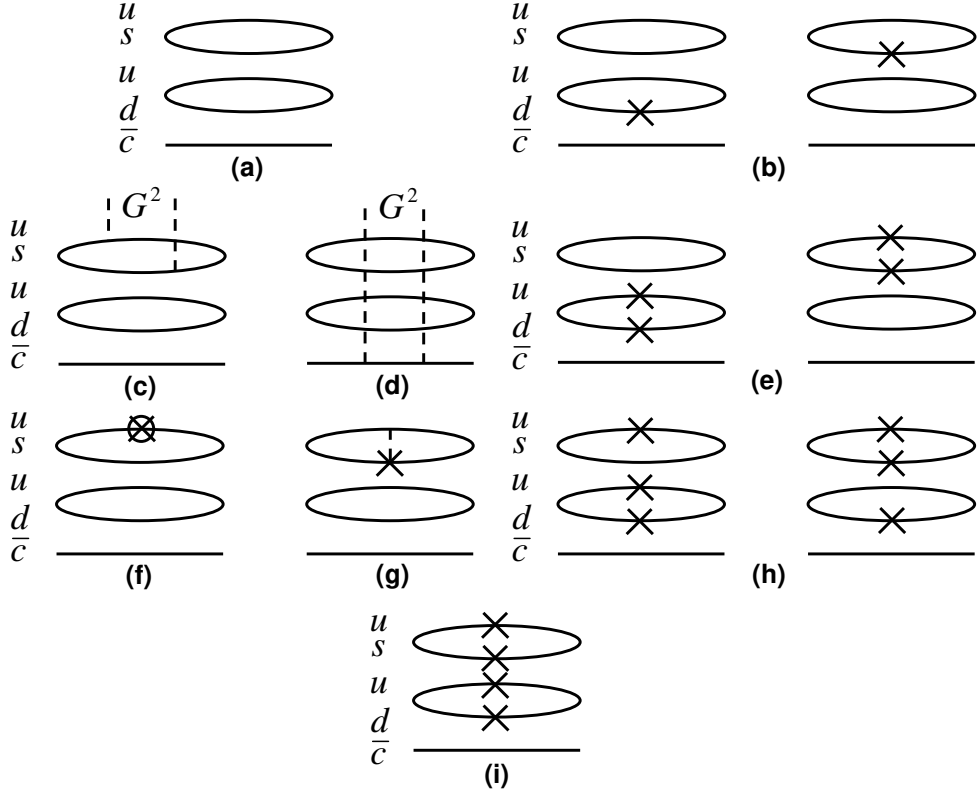, width=1.\hsize}
\caption{Schematic OPE diagrams for the currents $\Theta_{cs1}$ in
Eq.(\ref{ope-cs1}) and $\Theta_{cs2}$ in
Eq.(\ref{ope-cs2}). Each label corresponds to that in Eq.(\ref{ope-cs1}) or
Eq.(\ref{ope-cs2}). All the other notations in this figure are the
same as Fig.~\ref{fig1}.
} \label{fig2}
\end{figure}

\subsection{$\Theta_{cs1}$}

The OPE for this current is given as follows
\begin{eqnarray}
{1\over \pi} {\rm Im} \Pi^{(a)} (q^2) &=& \frac{1}{5\cdot 5!
\,2^{12} \, \pi^8} \int_{0}^{\Lambda} du \, \frac{1}{(1-u)^{5}}
\left\{ \fslash{q}(u-1)-m_c\right\} [-L(u)]^5
\ ,\nonumber \\
{1\over \pi} {\rm Im} \Pi^{(b)} (q^2) &=& {m_s(2
\langle\bar{q}q\rangle+\langle\bar{s}s\rangle)\over3!3!2^{8}\pi^6}\int_{0}^{\Lambda}du
\frac{1}{(1-u)^3}\left\{\fslash{q}(u-1)-m_c\right\}[-L(u)]^3
\ ,\nonumber \\
{1\over \pi} {\rm Im} \Pi^{(c)} (q^2) &=&
\frac{\langle\frac{\alpha_s}{\pi}G^2\rangle}{3!3!2^{10}\pi^6}
\int_{0}^{\Lambda}du\frac{1}{(1-u)^3}\left\{\fslash{q}(u-1)-mc\}\right[-L(u)]^3
\ ,\nonumber\\
{1\over \pi} {\rm Im} \Pi^{(d)} (q^2) &=&
-\frac{\langle\frac{\alpha_s}{\pi}G^2\rangle}{3\cdot3!5!2^{10}\pi^6}
\int_{0}^{\Lambda}du\frac{u^3}{(1-u)^5}\Big\{\fslash{q}3m_c^2(1-u)
\nonumber\\
&&+m_c(1-u)(3-5u)q^2+2um_c^3\Big\}[-L(u)]^2
\,\nonumber\\
{1\over \pi} {\rm Im} \Pi^{(e)}(q^2)&=&
\frac{(\langle\bar{q}q\rangle^2+\langle\bar{q}q\rangle\langle\bar{s}s\rangle)}{3\cdot2^{7}\pi^4}
\int_{0}^{\Lambda}du
\frac{1}{(1-u)^2}\left\{\fslash{q}(u-1)-m_c\}\right[-L(u)]^2
\ ,\nonumber\\
{1\over \pi} {\rm Im} \Pi^{(f)} (q^2) &=&
\frac{m_s\langle\bar{q}D^2q\rangle}{2^ {10}\pi^6}
\int_{0}^{\Lambda}du\frac{1}{(1-u)^2}\left\{\fslash{q}(u-1)-m_c\}\right[-L(u)]^2
,\nonumber\\
{1\over \pi} {\rm Im} \Pi^{(g)} (q^2) &=& \frac{m_s \langle \bar
{s}g\sigma\cdot G s \rangle}{3\cdot2^{11}\pi^6}
\int_{0}^{\Lambda}du\frac{1}{(1-u)^2}\left\{\fslash{q}(u-1)-m_c\}\right[-L(u)]^2
\ ,\nonumber \\
{1\over \pi} {\rm Im} \Pi^{(h)} (q^2) &=&
\frac{m_s(2\langle\bar{q}q\rangle^3+\langle\bar{q}q\rangle^2\langle\bar{s}s\rangle)}{9\cdot2^{4}\pi^2}
\int_{0}^{\Lambda}du \left\{\fslash{q}(u-1)-m_c\right\}
\ , \nonumber \\
{1\over \pi} {\rm Im} \Pi^{(i)} (q^2) &=&
\frac{\langle\bar{q}q\rangle^3\langle\bar{s}s\rangle}
{54}\left(\fslash{q}+m_c\right)\delta(q^2-m_c^2). \label{ope-cs1}
\end{eqnarray}

\begin{figure}
\centering \epsfig{file=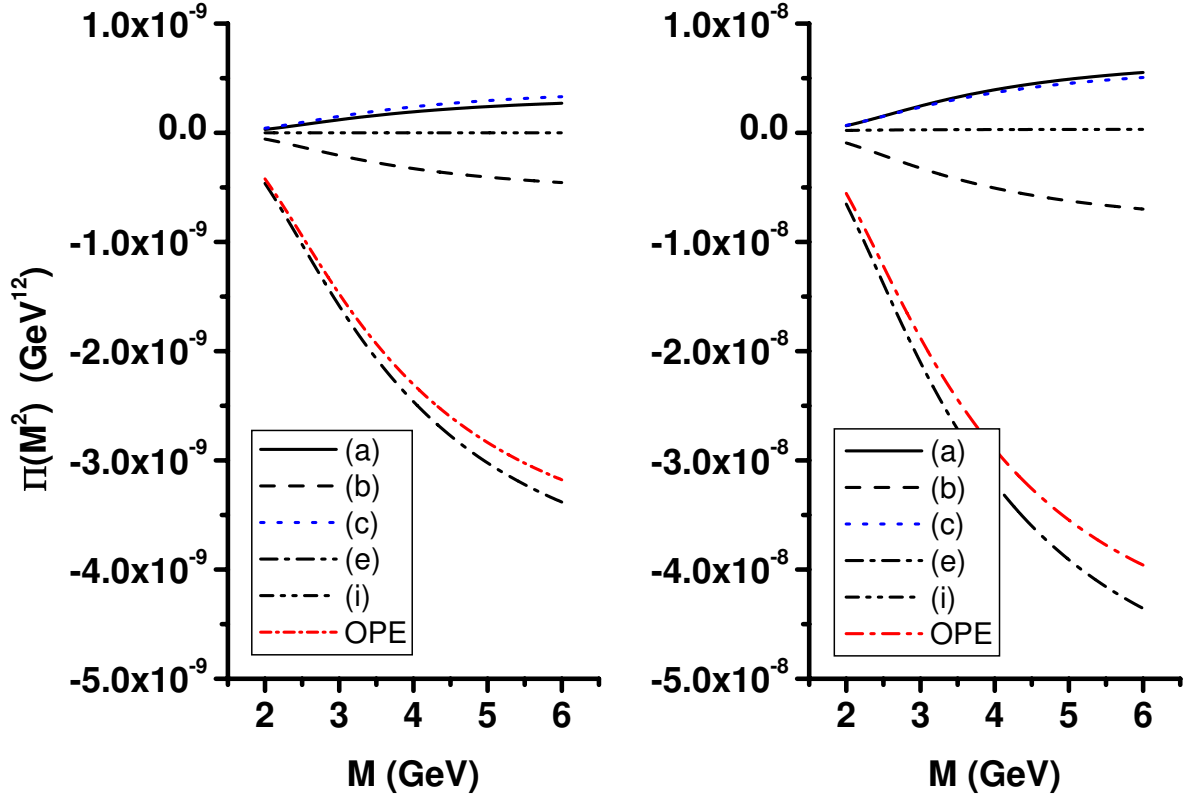, width=1.\hsize} \caption{
Similar figure as Fig.\ref{fig-borel-opec1} for the current $\Theta_{cs1}$
with $S_0=(3.3~{\rm GeV})^2$.
Here each label represents contribution from each term in
Eq.(\ref{ope-cs1}).}
\label{fig-borel-opecs1}
\end{figure}

Note here again that the superscripts correspond to the diagrams
shown in Fig.~\ref{fig2}. The dimension-5 condensate involving
$D^2$ is related to the quark-gluon condensate via
$\langle\bar{q}D^2q\rangle=\langle\bar{q}g\sigma\cdot
Gq\rangle/2$.  Similar relation holds for the  corresponding
strange-quark condensate. The correction to this relation is
proportional to square of the quark mass which should be very
small even for the strange quark. Fig. (\ref{fig-borel-opecs1})
represents the OPE as defined in Eq.(\ref{borel-opec1}) with the
imaginary part in Eq.(\ref{ope-cs1}).   We have only included a
few terms in the OPE to show how each term contributes differently
to the sum rule. As can be seen from the figure, the line denoted
as ``OPE'', which is sum of the power corrections only, are much
larger than the perturbative contribution.  Moreover, the gluon
condensate from diquarks is only a small fraction of the large
higher order correction.  This suggests that the OPE are not
convergent and it is very unlikely that the diquark correlation
will remain an important mechanism in this configuration.

\subsection{$\Theta_{cs2}$}

The OPE for this current is given as follows
\begin{eqnarray}
{1\over \pi} {\rm Im} \Pi^{(a)} (q^2) &=& \frac{1}{5\cdot 5!
\,2^{12} \, \pi^8} \int_{0}^{\Lambda} du \, \frac{1}{(1-u)^{5}}
\left\{ \fslash{q}(u-1)-m_c\right\} [-L(u)]^5
\ ,\nonumber \\
{1\over \pi} {\rm Im} \Pi^{(b)} (q^2) &=& {m_s(-2
\langle\bar{q}q\rangle+\langle\bar{s}s\rangle)\over3!3!2^{8}\pi^6}\int_{0}^{\Lambda}du
\frac{1}{(1-u)^3}\left\{\fslash{q}(u-1)-m_c\right\}[-L(u)]^3
\ ,\nonumber \\
{1\over \pi} {\rm Im} \Pi^{(c)} (q^2) &=&
\frac{\langle\frac{\alpha_s}{\pi}G^2\rangle}{3!3!2^{10}\pi^6}
\int_{0}^{\Lambda}du\frac{1}{(1-u)^3}\left\{\fslash{q}(u-1)-mc\}\right[-L(u)]^3
\ ,\nonumber\\
{1\over \pi} {\rm Im} \Pi^{(d)} (q^2) &=&
-\frac{\langle\frac{\alpha_s}{\pi}G^2\rangle}{3\cdot3!5!2^{10}\pi^6}
\int_{0}^{\Lambda}du\frac{u^3}{(1-u)^5}\left\{\fslash{q}3m_c^2(1-u)\right.
\nonumber\\
&&+\left.m_c(1-u)(3-5u)q^2+2um_c^3\}\right[-L(u)]^2
\,\nonumber\\
{1\over \pi} {\rm Im} \Pi^{(e)}(q^2)&=&
\frac{(\langle\bar{q}q\rangle^2-\langle\bar{q}q\rangle\langle\bar{s}s\rangle)}{3\cdot2^{7}\pi^4}
\int_{0}^{\Lambda}du
\frac{1}{(1-u)^2}\left\{\fslash{q}(u-1)-m_c\}\right[-L(u)]^2
\ ,\nonumber\\
{1\over \pi} {\rm Im} \Pi^{(f)} (q^2) &=&
\frac{-m_s\langle\bar{q}D^2q\rangle}{2^ {10}\pi^6}
\int_{0}^{\Lambda}du\frac{1}{(1-u)^2}\left\{\fslash{q}(u-1)-m_c\}\right[-L(u)]^2
,\nonumber\\
{1\over \pi} {\rm Im} \Pi^{(g)} (q^2) &=& \frac{m_s \langle \bar
{s}g\sigma\cdot G s \rangle}{3\cdot2^{11}\pi^6}
\int_{0}^{\Lambda}du\frac{1}{(1-u)^2}\left\{\fslash{q}(u-1)-m_c\}\right[-L(u)]^2
\ ,\nonumber \\
{1\over \pi} {\rm Im} \Pi^{(h)} (q^2) &=&
\frac{m_s(-2\langle\bar{q}q\rangle^3+\langle\bar{q}q\rangle^2\langle\bar{s}s\rangle)}
{9\cdot2^{4}\pi^2}\int_{0}^{\Lambda}du
\left\{\fslash{q}(u-1)-m_c\right\}
\ ,\nonumber \\
{1\over \pi} {\rm Im} \Pi^{(i)} (q^2) &=&
\frac{-\langle\bar{q}q\rangle^3\langle\bar{s}s\rangle}
{54}\left(\fslash{q}+m_c\right)\delta(q^2-m_c^2). \label{ope-cs2}
\end{eqnarray}

\begin{figure}
\centering \epsfig{file=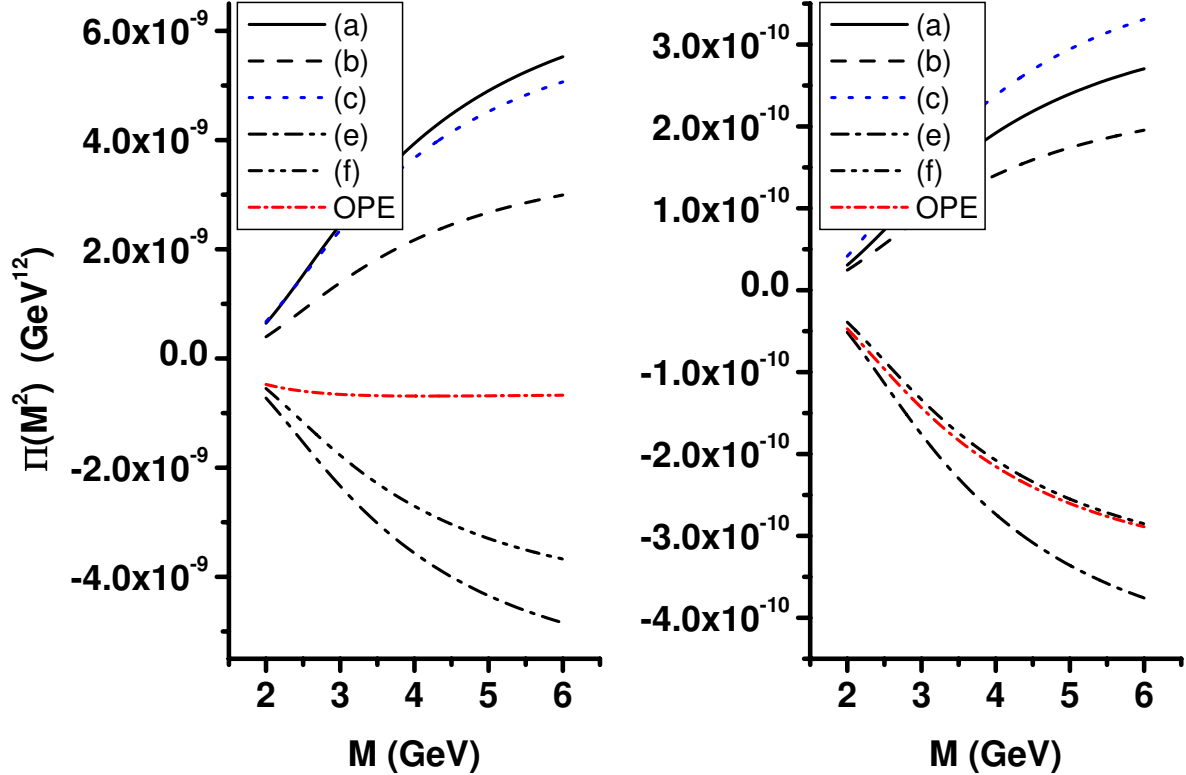, width=1.\hsize} \caption{
Similar figure as Fig.\ref{fig-borel-opec1} for the current $\Theta_{cs2}$
with $S_0=(3.3~{\rm GeV})^2$.
Here each label represents each term in Eq.(\ref{ope-cs2}).
} \label{fig-borel-opecs2}
\end{figure}

Again note that the OPE diagram for each label is
shown in Fig.~\ref{fig2}.
Fig. (\ref{fig-borel-opecs2}) represents the OPE as defined in
Eq.(\ref{borel-opec1}) with the imaginary part in
Eq.(\ref{ope-cs2}).  Again, we have only included a few terms
in the OPE to show a general trend of each contribution.
For the negative parity case, the OPE has large
contributions with alternating signs.   The situation is better
for the positive parity case, but again, the power corrections
alternate in signs.

{}From all the previous analysis on the OPE for the charmed
pentaquark with and without strangeness,  we find that the one
without strangeness with diquark structure are most reliable, and
are dominated by gluon condensate coming from diquark correlation.
It is interesting to note that this result is consistent with
the Skyrme model calculation which predicts a bound state
of pentaquarks in the nonstrange sector~\cite{OPM94}.
In the following, we will perform a more detailed analysis
with the stable structure well represented by the interpolating
current $\Theta_{c2}$.

\section{QCD sum rules and analysis}

\subsection{The couplings to the $DN$ continuum, $\lambda_{DN}$}

As discussed before, it is important to subtract out the
contribution from the $DN$ continuum.  For that, one needs to know
the coupling strength $\lambda_{DN}$. Here we determine this for
the currents without strange quarks, $\lambda_{DN,c2}$.   In the
case of $\Theta^+$ (1540)\cite{Lee04}, the soft-kaon theorem was used to
convert the external Kaon state, corresponding to the $D$ meson
states in Eq.(\ref{lambda1}) and Eq.(\ref{lambda2}), to a
commutation relation of the operator and the corresponding axial
charge.   The strength of the resulting five-quark operator with
an external nucleon state was then obtained from a separate
nucleon sum rule analysis with the same five-quark nucleon
current.   However, applying the soft $D$ meson limit will
obviously not work in the present case.

Instead, we determine the coupling strength directly from the sum
rule method. To do that, we eliminate contribution from the
low-lying pole by introducing the additional weight
$W(q^2)=q^2-m_\Theta^2$ in Eq.(\ref{sumrule1}).
We will take
$m_\Theta=3$ GeV,  and confirmed that changing it by $\pm 200$ MeV
will have less than 5 \% effect on the $\lambda_{DN}$ value.
This way of eliminating a certain pole is sometime used in QCD sum
rules~\cite{jin97,ko03}. Then, substituting the corresponding imaginary
parts, we find,
\begin{eqnarray}
|\lambda_{DN}|^2= { \int_{m_c^2}^{S_0} dq^2~e^{-q^2/M^2} (q^2-m_\Theta^2)
\frac{1}{\pi} {\rm Im} \Pi_i^{\rm ope}(q^2) \over
\int_{(m_N+m_D)^2}^{S_0} dq^2~e^{-q^2/M^2} (q^2-m_\Theta^2) \frac{1}{\pi} {\rm
Im} \Pi_i^{DN}(q^2)  },  \,\,\, (i=1,q)\label{lambdadn}
\end{eqnarray}
where the $i=1,q$ in Im$\Pi$ represent the part proportional to
$1$ or $\fslash{q}$ in the respective imaginary part, and
Im$\Pi^{DN}$ is the spectral density in Eq.(\ref{phen_dn1}) or in
Eq.(\ref{phen_dn2}) without the $|\lambda_{DN}|^2$.

\begin{figure}
\centering \epsfig{file=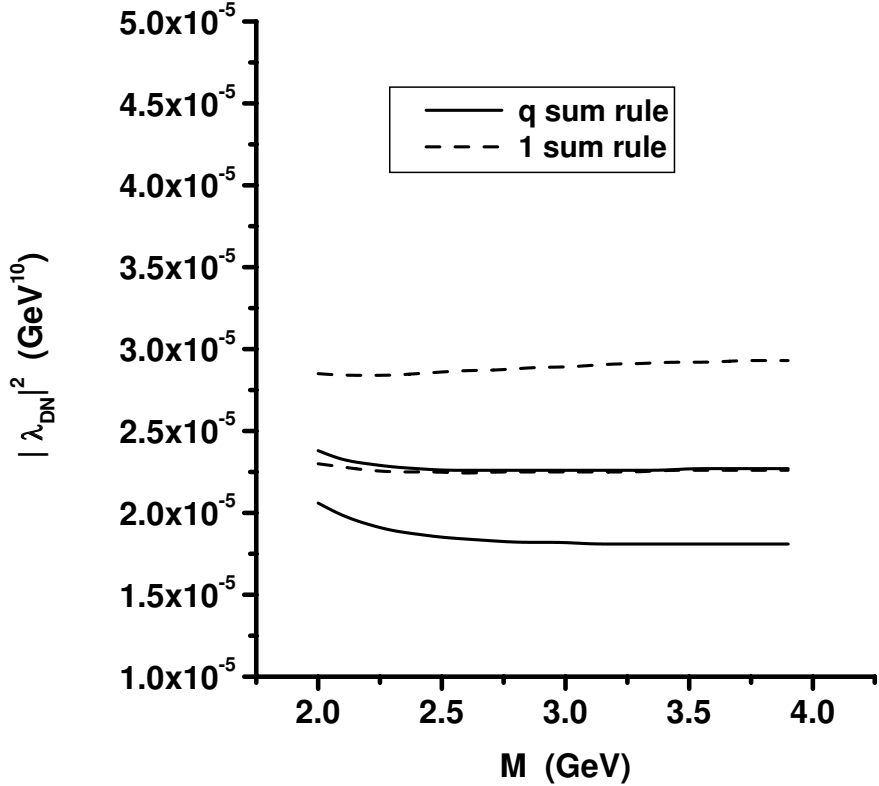, width=1.0\hsize} \caption{ The
$|\lambda_{DN,c2}|^2$ from the sum rule for $i=q$ (solid) line and
$i=1$ (dashed line).    The upper (lower) solid or dashed lines in
this case are for $S_0=(3.8~{\rm GeV})^2$ [$S_0=(3.7~{\rm
GeV})^2$]. } \label{lambda}
\end{figure}

Figure (\ref{lambda}) shows the plot of Eq.(\ref{lambdadn}). The
two dotted (solid) lines represent boundary curves with the least
Borel mass dependence for the $\lambda_{DN}$ from the  1 ($q$) sum
rules.   $\lambda_{DN}$ should not only be independent of the
Borel mass but also independent of the sum rule from which it is
obtained.   However, the results coming from either $i=q$ or $i=1$
sum rule differ slightly. Inspecting the OPE, one finds that the
contributions from higher dimensional operators are consecutively
suppressed for the $i=q$ sum rule, while that is not so for the
$i=1$ sum rule.  Therefore, the value from the former sum rule
should be more reliable.   Nonetheless, to allow for all
variations, we will choose the following range for the
$|\lambda_{DN}|^2 $ values,
\begin{eqnarray}
2 \times 10^{-5}{\rm GeV}^{10} < & |\lambda_{DN,c2}|^2 & < 3 \times
10^{-5} {\rm GeV}^{10}.
\end{eqnarray}
Similar attempts to determine $\lambda_{DN,c1}$ give vastly
different values from either $i=q$ or $i=1$ sum rules.  This
reflects the non-convergence of OPE from which one can not expect
a consistent result.

\subsection{Parity}

\begin{figure}
\centering \epsfig{file=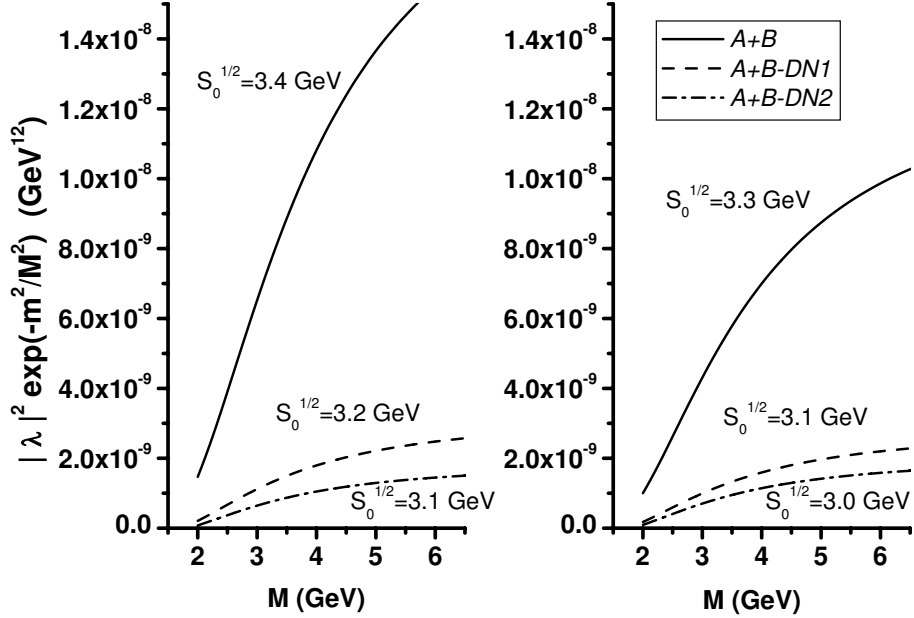, width=0.8\hsize}
\caption{The left figure shows the left-hand side of
Eq.(\ref{thetasum}) using $\Theta_{c2}$ for positive parity case
with $|\lambda_{DN,c2}|^2=2 \times 10^{-5}~ {\rm GeV}^{10}$
(dashed line) and $|\lambda_{DN,c2}|^2=3 \times 10^{-5}~ {\rm
GeV}^{10}$ (dot-dashed line). The solid line is when there is no
$DN$ continuum, $|\lambda_{DN,c2}|^2=0$. The right figure is
obtained with different threshold parameters.  See Eq. (8) for $A$
and $B$.  } \label{kimope-po}
\end{figure}

\begin{figure}
\centering \epsfig{file=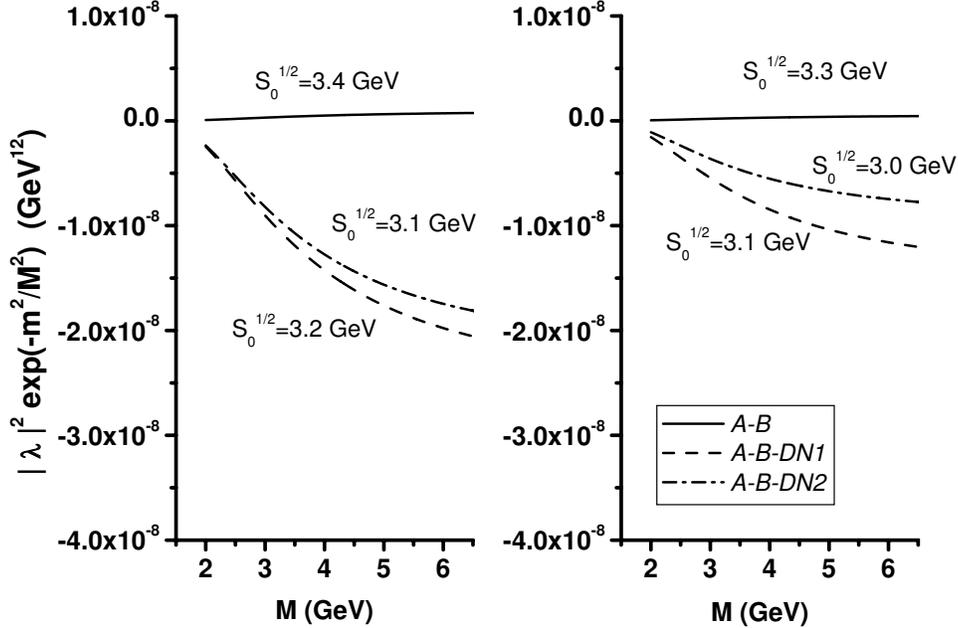, width=0.8\hsize}
\caption{The left figure shows the left-hand side of Eq.(\ref{thetasum})
using $\Theta_{c2}$ for negative parity case with $|\lambda_{DN,c2}|^2=2
\times 10^{-5}~ {\rm GeV}^{10}$ (dashed line) and
$|\lambda_{DN,c2}|^2=3 \times 10^{-5}~ {\rm GeV}^{10}$ (dot-dashed
line). The solid line is when there is no $DN$ continuum.
The right figure is obtained with different threshold parameters. }
\label{kimope-ne}
\end{figure}

We will now concentrate on the sum rule obtained from
$\Theta_{c2}$. Using the dispersion relation in
Eq.(\ref{sumrule2}) and the spectral density in
Eq.(\ref{phen-rho}), one finds the following sum rule,
\begin{eqnarray}
|\lambda_{\pm,c2}|^2 e^{-m_{\Theta \pm}^2/M^2 } =
\int_0^{\sqrt{s_0}} dq_0~ e^{-q_0^2/M^2} \bigg[ \rho_{\rm ope}^\pm
(q_0)-\rho_{DN}^\pm (q_0) \bigg]  \ . \label{thetasum}
\end{eqnarray}

 As can be seen from Fig.(\ref{kimope-po}), the left hand
side of Eq.(\ref{thetasum}) is positive for positive parity case.
For $|\lambda_{DN,c2}|^2=0$ (the solid lines), we have chosen the continuum
threshold $S_0^{1/2}$ to be 3.4 GeV and 3.3 GeV, which
gives the most stable pentaquark mass as we will show in the
next subsection. Similar method was used to obtain the continuum
thresholds when $|\lambda_{DN,c2}|^2 \neq0$.    However,
Fig.(\ref{kimope-ne}) shows that the corresponding sum rule  is
negative for the negative parity case, suggesting that there can
not be any negative parity state.  This result also confirms the
non-convergence of the OPE for the negative parity case, from
which a consistent result can not be obtained. This can also be
expected from the constituent quark picture. The two diquarks in
$\Theta_{c2}$ current have opposite parities and, when they are
combined with the antiquark, the configuration should be dominated
by the positive-parity part in the nonrelativistic limit.

\subsection{Mass}

\begin{figure}
\centering \epsfig{file=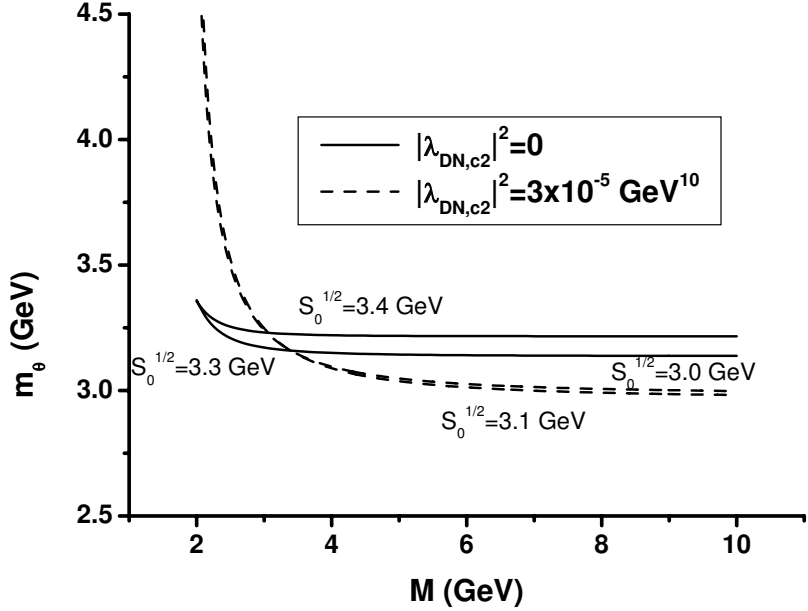, width=0.8\hsize}
\caption{The mass obtained by taking the square root of the
inverse ratio between left hand side of Eq.(\ref{thetasum}) and
its derivative with respect to $M^2$ using  $\Theta_{c2}$.   }
\label{mass-lambda}
\end{figure}

The sum rule for the $\Theta_c$ mass is obtained by taking the
derivative of Eq.(\ref{thetasum}) with respect to $1/M^2$.  The
solid  and dashed lines in Fig. (\ref{mass-lambda}) represent the
mass for two different $\lambda_{DN}$ values.  The threshold
parameters were obtained to give the most stable mass within the
Borel window plotted.  One notes that the inclusion of the
coupling to the $DN$ continuum states, the mass reduces to smaller
values to below 3 GeV.  The curve with $\lambda_{DN,c2}=2\times
10^{-5}$ GeV${}^{10}$ lies between the solid and dashed lines in
Fig. (\ref{mass-lambda}).   This suggests the possibility that the
heavy pentaquark might actually be bound; namely, lies below the
$DN$ threshold. This is consistent with the constituent quark
model picture, where one expects the diquark correlation to be
more dominant than that of the quark-antiquark correlation as the
participating antiquark becomes heavy. However, if this was the
case, its existence can only be measured through its weak decay.

\section{summary}

We have performed the OPE and QCD sum rule analysis for heavy
pentaquark with and without strangeness with two different current
each.   We find that the OPE is convergent only for the
non-strange pentaquark with diquark structure.   The OPE for this
structure is dominated by gluon condensate coming from diquark,
which non-perturbatively represents their strong correlation.  We
find that  the heavy pentaquark without strangeness has positive
parity as reported earlier\cite{hung04} and that its mass lies
below 3 GeV, when the $DN$ irreducible contribution is explicitly
including in the phenomenological side of the sum rule.  The
picture that we described here does not work so well in the light
pentaquark $\Theta^+$, as the OPE are highly divergent\cite{MN04}
as can be seen in the picture of the OPE in the original sum rule
paper for the light pentaquark state\cite{SDO03}.

\acknowledgments

We are grateful to A. Hosaka and Fl. Stancu for fruitful
discussions. The work of S.H.L was supported by Korea research
foundation under grant number C00116.
The work of Y. S was supported by the Scientific and
Technological Research Council of Turkey.

\end{document}